\documentclass[letterpaper, 10 pt, conference]{ieeeconf}  

\IEEEoverridecommandlockouts                              

\usepackage{graphics} 
\usepackage{epsfig} 
\usepackage{mathtools}
\usepackage{amsmath} 
\usepackage{amssymb}  
\usepackage{algpseudocode}
\usepackage{algorithm}
\usepackage{todonotes}
\usepackage{xcolor}
\usepackage{graphicx}
\usepackage{caption,subcaption}

\usepackage{multirow}
\usepackage{mathrsfs}
\usepackage{ctable} 
\usepackage{url}
\usepackage{array,booktabs,ragged2e}
\usepackage[english]{babel}
\usepackage{gensymb}
\usepackage{hhline}
\usepackage{hyperref}
\usepackage{amsmath}
\usepackage{ctable} 
\usepackage{array}

\newtheorem{problem}{Problem}
\newtheorem{lemma}{Lemma}

\newtheorem{proposition}{Proposition}
\newtheorem{remark}{Remark}
\newtheorem{definition}{Definition}

\renewcommand{\r}{{\mathbf r}}

\newcommand{\A}{\mathcal{A}}
\newcommand{\R}{\mathbb{R}}

\newcommand{\x}{\mathbf{x}}
\newcommand{\y}{\mathbf{y}}

\renewcommand{\u}{{\mathbf u}}

\renewcommand{\A}{{\mathcal A}}
\newcommand{\0}{{\mathbf 0}}

\newcommand{\llambda}{{\boldsymbol \lambda}}

\usepackage{tikz}

\newcommand\copyrighttext{%
  \footnotesize \textcopyright 2024 IEEE.  Personal use of this material is permitted.  Permission from IEEE must be obtained for all other uses, in any current or future media, including reprinting/republishing this material for advertising or promotional purposes, creating new collective works, for resale or redistribution to servers or lists, or reuse of any copyrighted component of this work in other works.}
\newcommand\copyrightnotice{%
\begin{tikzpicture}[remember picture,overlay]
\node[anchor=south,yshift=10pt] at (current page.south) {\fbox{\parbox{\dimexpr\textwidth-\fboxsep-\fboxrule\relax}{\copyrighttext}}};
\end{tikzpicture}%
}

\title{\LARGE \bf
Estimation of Constraint Admissible Invariant Set \\ with Neural Lyapunov Function}

\author{Dabin Kim and H. Jin Kim
\thanks{This research was supported by ITRC (IITP) (IITP-2024-RS-2024-00437268).}
\thanks{The authors are with the Department of Aerospace Engineering, Seoul National University, Seoul 08826, South Korea (e-mail: dabin404@snu.ac.kr, hjinkim@snu.ac.kr, corresponding author: H. Jin Kim).}
}

\begin{document}
\maketitle
\copyrightnotice
\thispagestyle{empty}
\pagestyle{empty}

\vspace{-3.5mm}
\begin{abstract}
Constraint admissible positively invariant (CAPI) sets play a pivotal role in ensuring safety in control and planning applications, such as the recursive feasibility guarantee of explicit reference governor and model predictive control. However, existing methods for finding CAPI sets for nonlinear systems are often limited to single equilibria or specific system dynamics. This limitation underscores the necessity for a method to construct a CAPI set for general reference tracking control and a broader range of systems. In this work, we leverage recent advancements in learning-based methods to derive Lyapunov functions, particularly focusing on those with piecewise-affine activation functions. Previous attempts to find an invariant set with the piecewise-affine neural Lyapunov function have focused on the estimation of the region of attraction with mixed integer programs. We propose a methodology to determine the maximal CAPI set for any reference with the neural Lyapunov function by transforming the problem into multiple linear programs. Additionally, to enhance applicability in real-time control scenarios, we introduce a learning-based approach to train the estimator, which infers the CAPI set from a given reference. 
The proposed approach is validated with multiple simulations to show that it can generate a valid CAPI set with the given neural Lyapunov functions for any reference. We also employ the proposed CAPI set estimation method in the explicit reference governor and demonstrate its effectiveness for constrained control.
 
\end{abstract}

\section{Introduction}
Positively invariant sets, also referred to as invariant sets in the rest of the paper, are subsets of the state space which once the state enters, it remains within them indefinitely. Such sets find extensive application in constrained control, particularly in contexts demanding high safety standards, such as obstacle avoidance in automated vehicle systems \cite{danielson2016path}. Explicit reference governors \cite{garone2015explicit} leverage invariant sets to calculate safety margins for a given reference, and model predictive control employs the invariant set as terminal constraints to ensure recursive feasibility \cite{allgower2012nonlinear}.

Finding a systematic methodology for constructing invariant sets has been the subject of studies for different classes of dynamical systems, including linear hybrid systems \cite{haimovich2010componentwise}, polynomial systems \cite{oustry2019inner}, and piece-wise affine systems \cite{samanipour2024invariant}.
However, these works address the characterization of domains of attraction with respect to single equilibrium states. But for reference tracking control, there is a necessity for establishing invariant sets surrounding any equilibrium point. Also, in order to be used in constrained control or safe planning applications, it is required to identify constraint admissible positive invariant (CAPI) sets, which are invariant subsets of the admissible set. 


Recently, several papers aimed at finding the CAPI set systematically for the application of constrained control or safe motion planning. However, they are limited for polynomial systems using numerically sensitive sum-of-squares (SoS) optimization \cite{cotorruelo2021reference} or only handle specific dynamics such as manipulator \cite{brandt2023safe} and satellite \cite{danielson2023rapid}. To fill the gap, the interest of the paper is to estimate the CAPI set for more general nonlinear systems with improved computational efficiency and stability. In this paper, similar to previous work \cite{brandt2023safe}, we compute the CAPI set as a sublevel set of the Lyapunov function of an asymptotically stable system. The sublevel set of the Lyapunov function inherently satisfies positive invariance. Thus, finding the maximal CAPI set can be transformed into finding the maximal level of the Lyapunov function, which makes the sublevel set admissible.

Using the Lyapunov sublevel set as a CAPI set requires a pre-defined Lyapunov function, but synthesizing such functions for general nonlinear systems is challenging \cite{giesl2015review}.  Following the  growing interest in learning-based methods in control, there have been multiple studies to find a Lyapunov function parameterized with neural network \cite{dai2021lyapunov} \cite{chang2019neural} \cite{wu2023neural}. These neural Lyapunov constructing methods have been demonstrated Lyapunov functions for various types of nonlinear systems with enlarged region of attraction compared to the Lyapunov functions from LQR or SoS methods.  
 
The paper proposes to estimate the maximal CAPI set by utilizing the neural Lyapunov function with continuous piecewise-affine (PWA) activation functions.
The invariant set estimation with the neural Lyapunov function from previous works \cite{dai2021lyapunov} \cite{wu2023neural} is limited to the estimation of the region of attraction and done with solving the computationally demanding mixed-integer linear program.
The major contribution of the paper compared to previous works is that it enables estimation of the admissible invariant set at any reference from the neural Lyapunov function with improved computational efficiency.
 The main idea is to transform the problem of the computation of a CAPI set into a set of multiple linear programs that can be solved efficiently. By utilizing the property that a neural network with PWA activation functions can be expressed as a continuous PWA function, we divide the process of finding the maximal admissible Lyapunov level into the optimization problems for each partition where the Lyapunov function is affine. 

 

Additionally, in order to expand the usage of the proposed method in control methods that require real-time inference of the CAPI set, such as explicit reference governor, we suggest a data-driven method to train an estimator, which is a neural network whose output is the maximal admissible Lyapunov level. The training procedure is based on a counterexample-guided inductive synthesis (CEGIS), which is composed of a verifier and learner, where the verifier checks that the estimator meets safe requirements and generates a counterexample that is later used by the learner to train the estimator.  



The contributions of our work can be summarized as:
\begin{itemize}
    \item We propose a computation method for the CAPI set for asymptotically stable nonlinear systems with a neural Lyapunov function. 
    \item We suggest an application for neural Lyapunov functions for the control of safety-critical systems.
    \item We extend the usage of the proposed computation method to real-time online control by designing the data-driven estimator of the CAPI set.
    \item We validated the proposed method with numerical studies and applied it to the explicit reference governor for constrained control. 
\end{itemize}


\section{Background \& Problem Description}
\subsection{Reference Dependent Lyapunov Function}
For the state space \( D \subset \mathbb{R}^{n} \), the state vector $\x \in D$, the domain of reference \( R \subset \mathbb{R}^{n_{r}} \), the reference vector $\r \in R$, and the discrete-time dynamical system \( f \), the reference-dependent Lyapunov function (RLDF) \( V:D\times R \rightarrow \mathbb{R} \) \cite{cotorruelo2021reference} is a continuous function satisfying the following three conditions:
\begin{subequations}     \label{eq:lyap_3}
\begin{align}
    V(\bar{\mathbf{x}}_{\mathbf{r}}, \mathbf{r}) &= 0, \label{eq:lyap_zero_condition} \\
    V(\mathbf{x},\mathbf{r}) &> 0, \quad \forall \mathbf{x} \in D \setminus \{\bar{\mathbf{x}}_{\mathbf{r}}\}, \\
    V(f(\x),\r)-V(\x,\r) &\leq 0, \quad \forall \mathbf{x} \in D \setminus \{\bar{\mathbf{x}}_{\mathbf{r}}\},
\end{align}
\end{subequations}
where $\bar{\x}_{\r}$ is an equilibrium state corresponding to the reference $\r$. 
If there exists a valid RLDF, the system \( f \) is ensured to have local stability in the state space \( D \) \cite{khalil2002control}. Asymptotic stability is achieved if the strict inequality of condition \eqref{eq:lyap_3} is met. The sublevel set of an asymptotic RLDF with the level \( \Gamma \), \( \{\mathbf{x} \in D: V(\mathbf{x},\mathbf{r}) \leq \Gamma, \Gamma >0 \} \), is positively invariant.

In general literature on constructing Lyapunov function is focused on (reference-independent) Lyapunov function for a single equilibrium at origin. To use the previous Lyapunov function synthesis method in our work, we make an assumption regarding the dynamics of the system. 
\begin{definition}[Reference Symmetry \cite{jang2024safe}]
 The closed-loop discrete-time dynamics $f_{cl}$ is \textit{reference symmetric} if there exists a matrix $E\in \R^{n\times n_{r}}$ such that $f_{cl}(\x,\r) = f_{cl}(\x-E\r,\0_{n_{r}})+E\r$ for any reference $\r \in \R^{r}$, where $\0_{n_{r}}\in \R^{n_{r}}$ is a zero vector. Denote $\bar{\x}_{\r}=E \r$ as an equilibrium state corresponding to the refernce $\r$. 
\end{definition}

The reference symmetry is a common property satisfied by many dynamical systems such as the mobile robot systems, which are symmetric with respect to their positions. 
\begin{lemma} \label{lemma:rdlf}
    Let $V'(\x):\R^{n}\rightarrow\R$ is a (reference independent) Lyapunov function which is equivalent to a reference-dependent Lyapunov function at zero reference $V(\x,\0_{n_{r}})$, and it is assumed to satisfy the reference-independent version of \eqref{eq:lyap_3}, 1) $V'(\0_{n})=0$, 2) $V'(\x)>0, \forall x \in D \setminus \{\0_{n}\}$, 3) $V'(f(\x))-V'(\x)\leq 0, \forall x \in D \setminus \{\0_{n}\}$.
    If the system is reference symmetric, the reference-dependent Lyapunov function can be constructed from a reference-independent Lyapunov function as 
    $V(\x,\r)=V'(\x-E \r)=V'(\x-\bar{\x}_{\r})$.
\end{lemma}
\begin{proof}
We need to check whether the Lyapunov candidate $V(\x,\r)=V'(\x-E \r)$ satisfies the conditions \eqref{eq:lyap_3}. The first and second conditions of \eqref{eq:lyap_3} are automatically satisfied. The third condition is satisfied as 
$V(f(\x,\r),\r)-V(\x,\r)= V'(f(\x,\r)-E\r)-V'(\x-E\r)= V'(f(\y))-V'(\y) \leq 0$
where $\y=\x-E\r$. Thus, $V(\x,\r)$ is a valid RLDF. 
\end{proof}

Thus, we do not need to construct an RLDF in dimension of $n+r$; instead, we can utilize a Lyapunov function derived from the equilibrium at origin and reconstruct it using Lemma \ref{lemma:rdlf}.


\subsection{Piecewise-Affine Function}

A piecewise-affine (PWA) function \( f_{\text{PWA}} \) is an affine function over a set of finite partitions \( \mathcal{P} \), which are convex polytopes, expressed as
\begin{align}
    f_{\text{PWA}}(\mathbf{x}) = C_{P} \mathbf{x} + \mathbf{d}_{P}, \quad \forall \mathbf{x} \in P, \quad \forall P \in \mathcal{P}, \label{eq:pwa}
\end{align}
where \( C_{P} \) and \( \mathbf{d}_{P} \) are coefficients corresponding to the partition \( P \).
If the PWA function is defined on a domain \( D \), there exists a partition set \( \mathcal{P} = \{ P_{1}, \cdots, P_{N} \} \) such that \( \cup_{i=1}^{N} P_{i} = D \) and \( P_{i} \cap P_{j} = \emptyset \) for \( i \neq j \).



\begin{figure}[t]
    \centering
    \vspace{2mm}
        \begin{tabular}[t]{cc}
		  \subfloat[]{\includegraphics[height=0.40\linewidth]{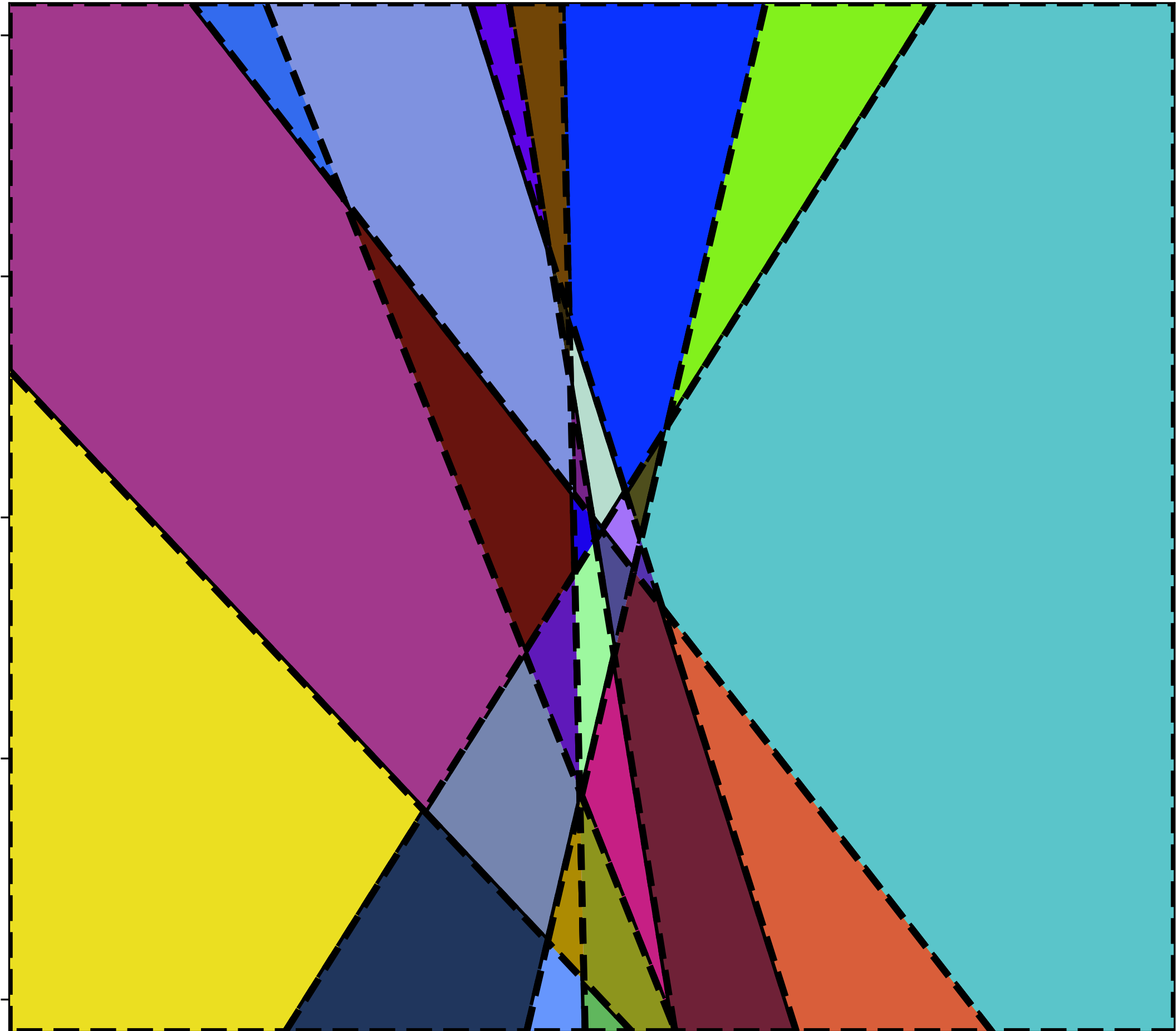}}
        & \hspace{2mm}
		  \subfloat[]{\includegraphics[height=0.405\linewidth]{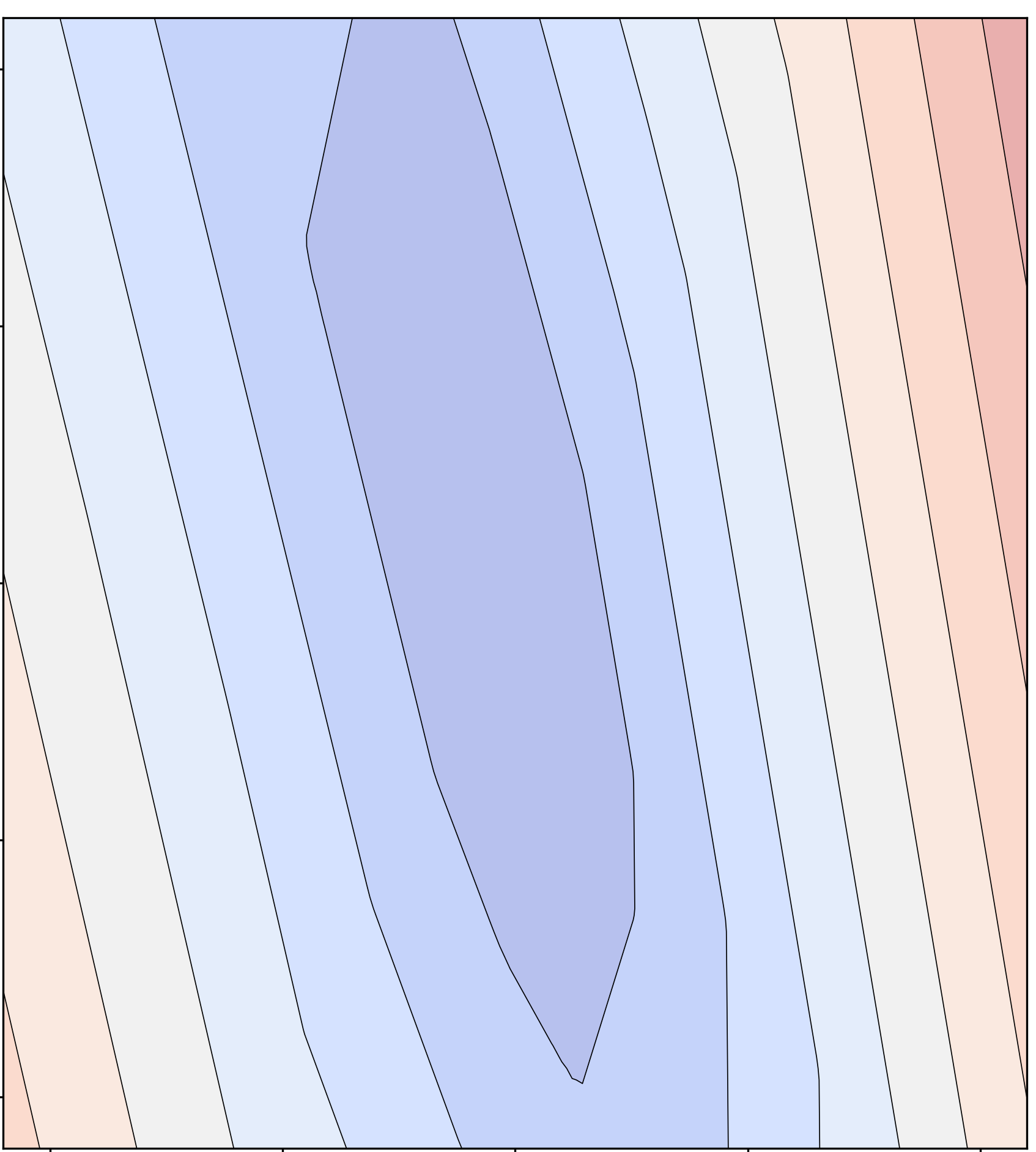}}
    \end{tabular}
    \caption{(a) Partitions and (b) contour plot of piecewise-affine neural Lyapunov function of inverted pendulum dynamics. Details can be found in Sec. \ref{sec:eval}-A.}
    \label{fig:nn_lyapunov}
\end{figure}

\subsection{Piecewise-Affine Neural Network}
The piecewise-affine neural network (NN) is a feed-forward neural network with a piecewise-affine activation function (e.g., ReLU, leaky-ReLU, PReLU) for hidden layers. In this paper, we focus on the activation function \( \sigma\) with a single breakpoint at zero, such as ReLU, although the proposed method can also accommodate activation functions with multiple breakpoints.

For the \( L \)-layer NN, the input-output relationship of each layer is given by
\begin{align}
    \mathbf{z}^{(l)} &= \sigma( F^{(l)}\mathbf{z}^{(l-1)} + \mathbf{b}^{(l)}), \quad l \in \{1, 2, \ldots, L-1\}, \label{eq:nn_io} \\
    \mathbf{z}^{(L)} &= F^{(L)}\mathbf{z}^{(L-1)} + \mathbf{b}^{(L)},
\end{align}
where \( \mathbf{z}^{(i)} \) is the output vector of the \( i \)-th layer and \( \mathbf{z}^{(0)} = \mathbf{x} \). $F^{(i)}\in \R^{w_{i}\times w_{i-1}}$ and $\mathbf{b}^{(i)}\in \R^{l_{i}}$ are the weight matrix and the bias vector for the  \( i \)-th layer, where $w_{i}$ denotes the number of neurons in the $i$-th layer. Using the homogeneous form, \eqref{eq:nn_io} can be represented as
\begin{align}
    \begin{bmatrix}
        \mathbf{z}^{(l)} \\ \sigma(1)
    \end{bmatrix}
    = \sigma \left( \Theta^{(l)}
    \begin{bmatrix}
        \mathbf{z}^{(l-1)} \\ 1
    \end{bmatrix}
    \right), 
    \quad 
    \Theta^{(l)} = \begin{bmatrix}
    F^{(l)} & \mathbf{b}^{(l)}  \\ \mathbf{0}^{T} & 1   
    \end{bmatrix}. \label{eq:nn_homo}
\end{align}

The neuron activation is defined as an indicator of whether the pre-activation value of the neuron is positive or not. We define the activation pattern at layer \(i\) as a vector of binary variables \(\boldsymbol{\lambda}^{(i)}=[\lambda^{(i)}_{1}, \cdots,\lambda^{(i)}_{W_{i}}]^{T}\), where
\begin{align}
    \lambda_{j}^{(i)} = \begin{cases}
    1 & \text{if } z^{(i)}_{\text{pre},j}(\mathbf{x}) > 0 \\
    0 & \text{if } z^{(i)}_{\text{pre},j}(\mathbf{x}) \leq 0
    \end{cases}, \quad j\in \{1,\cdots,w_{i}\}, \label{eq:ap}
\end{align}
and $z_{pre,j}^{(i)}(\x)$ is the pre-activation value at the $j$-th neuron of layer $i$.
The activation pattern of the whole network is a tuple of each hidden layer's activation pattern \(\boldsymbol{\lambda}=(\boldsymbol{\lambda}^{(1)},\cdots,\boldsymbol{\lambda}^{(L-1)})\). The matrix for the activation pattern (AP) of layer \(l\) from input \(\mathbf{x}\) is defined as \(\Lambda^{(l)}=\text{diag}([\boldsymbol{\lambda}^{(l)}(\mathbf{x})^{T}, 1]^{T})\) \footnote{$\text{diag}(\x)\in \R^{n\times n}$ is the diagonal matrix where its diagonal entries are elements of the vector $\x\in\R^{n}$.}. 

From the homogeneous form and \eqref{eq:ap}, the pre-activated value of the output vector of the hidden layer $l$ with respect to the input is formalized as
\begin{align} \label{eq:nn_hidden}
    z_{\text{pre},j}^{(l)}=\theta_{i}^{(l)} \left(\prod_{k=1}^{l-1} \Lambda^{(k)}(\mathbf{x}) \Theta^{(k)}  \right) \begin{bmatrix}
        \mathbf{x} \\ 1
    \end{bmatrix},
\end{align}
where $\theta_{i}^{(l)}$ is $i$-th row vector of $\Theta^{(l)}$. The output of the NN is expressed as 
\begin{align}
    \mathbf{y} = \Theta^{(L)}  \left(\prod_{k=1}^{L-1} \Lambda^{(k)}(\mathbf{x}) \Theta^{(k)}  \right) \begin{bmatrix}
        \mathbf{x} \\ 1
    \end{bmatrix}. \label{eq:nn_output}
\end{align}
which is a piecewise-affine function \cite{vincent2021reachable}, and can be represented in the form of \eqref{eq:pwa}. Since the AP is unique for each partition, with a slight abuse of notation, we denote the AP of partition \(P\) as \(\boldsymbol{\lambda}(P)\). Example of the neural Lyapunov function parameterized with PWA NN for the inverted pendulum dynamics is given in Fig. \ref{fig:nn_lyapunov}.


\subsection{Problem Description}
The nonlinear discrete-time system $f$ and a stabilizing control policy $\u=\pi(\x,\r)$ with a reference $\r \in R$ forms closed-loop system $f_{cl}$ as 
\begin{align}	
\x(t+1) &= f(\x(t),\u(t)) = f(\x(t),\pi(\x(t),\r)) \nonumber
 \\  &= f_{cl}(\x(t),\r). \label{eq:cl_dyn}
\end{align}
Assume that the corresponding neural Lyapunov function $V_{NN}$ which is parameterized by PWA NN is given. 

Let the admissible set $\mathcal{A} \subset D$ be expressed as a set of $\mathcal{A}=\{\x:c_{i}(\x) \leq 0, i=1,\cdots,n_{c}\}$, where all constraint functions $c_{i}(\x)$ are continuous PWA functions. Thus, they can be represented in the form of \eqref{eq:pwa}. Note that specifying the constraint function to a continuous PWA function is not restrictive since it includes functions parameterized with PWA NN.  

The objective of the paper is to find a maximal CAPI set with the given reference $\r$ from the neural Lyapunov function  $V_{NN}$. The CAPI set induced from the Lyapunov function $V_{NN}(\x,\r)$ would be the sublevel set of the Lyapunov level $\Gamma$  $A_{\Gamma}(\r)=\{\x\in D: V_{NN}(\x,\r)\leq \Gamma\}$ which is subset of the admissible set $\A$ as $A_{\Gamma}(\r)\subset \A$. The maximal CAPI set from the Lyapunov function can be found by solving the following optimization problem,  
\begin{align}
    \Gamma^{*}(\r) = \max \Gamma \hspace{3mm} \text{s.t. } A_{\Gamma}(\r) \subset \A,
\end{align}
which results in the maximal admissible Lyapunov level $\Gamma^{*}(\r)$ from given reference $\r\in R$. 


\begin{remark}
When dealing with multiple constraints $c_{1},\cdots, c_{n_{c}}$, the maximal admissible Lyapunov level for the whole constraints $\Gamma^{*}$ is  the minimum value of the maximal admissible Lyapunov level across all constraints as $\Gamma^{*}=\min\{\Gamma^{*}_{c_{1}}, \cdots, \Gamma^{*}_{c_{n_{c}}}\}$, where $\Gamma^{*}_{c_{i}}$ is the maximal admissible Lyapunov level for the constraint $c_{i}$. For the sake of brevity in explanation, we will henceforth consider a single constraint function $c(\x)$.
\end{remark}
\begin{remark}
We do not explicitly consider the input constraint since it is assumed to be achieved by the stabilizing controller $\u=\pi(\x,\r)$, such as by saturating control. 
\end{remark}

%

\section{Computation of Maximal Admissible Lyapunov Sublevel Set}
 



\subsection{Computation of the Maximal Admissible Lyapunov Level}


For the computation of the maximal level of the PWA NN Lyapunov function within the specified set, previous works \cite{dai2021lyapunov} and \cite{wu2023neural} utilized mixed-integer linear programming (MILP) to compute the region of attraction of the system. However, we propose that by leveraging the properties of the PWA NN, we can convert the problem into a series of small LPs. 

\begin{problem}
For the PWA Lyapunov function $V$ with the partition set $\mathcal{P}_{V}$ and the PWA constraint function $c(\mathbf{x})$ with the partition set $\mathcal{P}_{c}$, the maximal admissible Lyapunov level for the reference $\mathbf{r} \in  R$ can be found by solving the following optimization problem:
\begin{align}
    \Gamma^{*}(\mathbf{r}) = \min_{P_{c}\in \mathcal{P}_{c}, P_{V}\in\mathcal{P}_{V}} \Gamma_{P_{V},P_{c}}(\mathbf{r}), \label{eq:opt_problem}
\end{align}
where $\Gamma_{P_{V},P_{c}}$ denotes the solution of the following LP designed for a single partition pair $(P_{V}, P_{c})$:
\begin{subequations} \label{prob:single_lp}
\begin{align}
    \Gamma_{P_{V},P_{c}}(\r)= \min &\hspace{2mm}  C_{P_{V}}(\x-\bar{\x}_{\r})+d_{P_{V}} \\
    \text{s.t.} \hspace{2mm} &\x \in P_{V} \cap P_{C}, \\
     \hspace{6mm} &C_{P_{c}}\x+d_{P_{c}} = 0.
\end{align}
\end{subequations}
\end{problem}



\begin{proposition} \label{prop:optimal}
Solution of the Problem \eqref{eq:opt_problem} is the maximal admissible Lyapunov level from the given PWA Lyapunov function and constraint function, where its sublevel set is a CAPI set. 
\end{proposition}
\begin{proof}
For a given $\r \in R$, the condition of the admissible Lyapunov level $\Gamma(\r)$,
$\{ \x \in D : V(\x,\r)\leq \Gamma(\r)\} \subset \mathcal{A}$, can be restated as $\{c(\x,\r)=0, V(\x,\r)<\Gamma(\r)\}=\emptyset$. Thus,  $\Gamma^{*}(\r)$ is the minimum Lyapunov value at the boundary of the admissible set $c(\x,\r)=0$. Since $V$ and $c$ are continuous PWA functions, it can be found by solving \eqref{eq:opt_problem}. 
\end{proof}

Directly solving the optimization problem \eqref{eq:opt_problem} is not feasible due to the absence of an explicit representation of the PWA NN Lyapunov function in the form of \eqref{eq:pwa}. Consequently, it is necessary to convert $V_{NN}$ into a partition set and coefficient set. In the subsequent subsections, we elaborate on the computation of the partition and coefficient sets from the PWA NN Lyapunov function and introduce the algorithm aimed at addressing the optimization problem \eqref{eq:opt_problem}. 

\subsection{Partition Tree Construction}

\begin{figure}[t]
    \centering
    \includegraphics[width=1.0\linewidth]{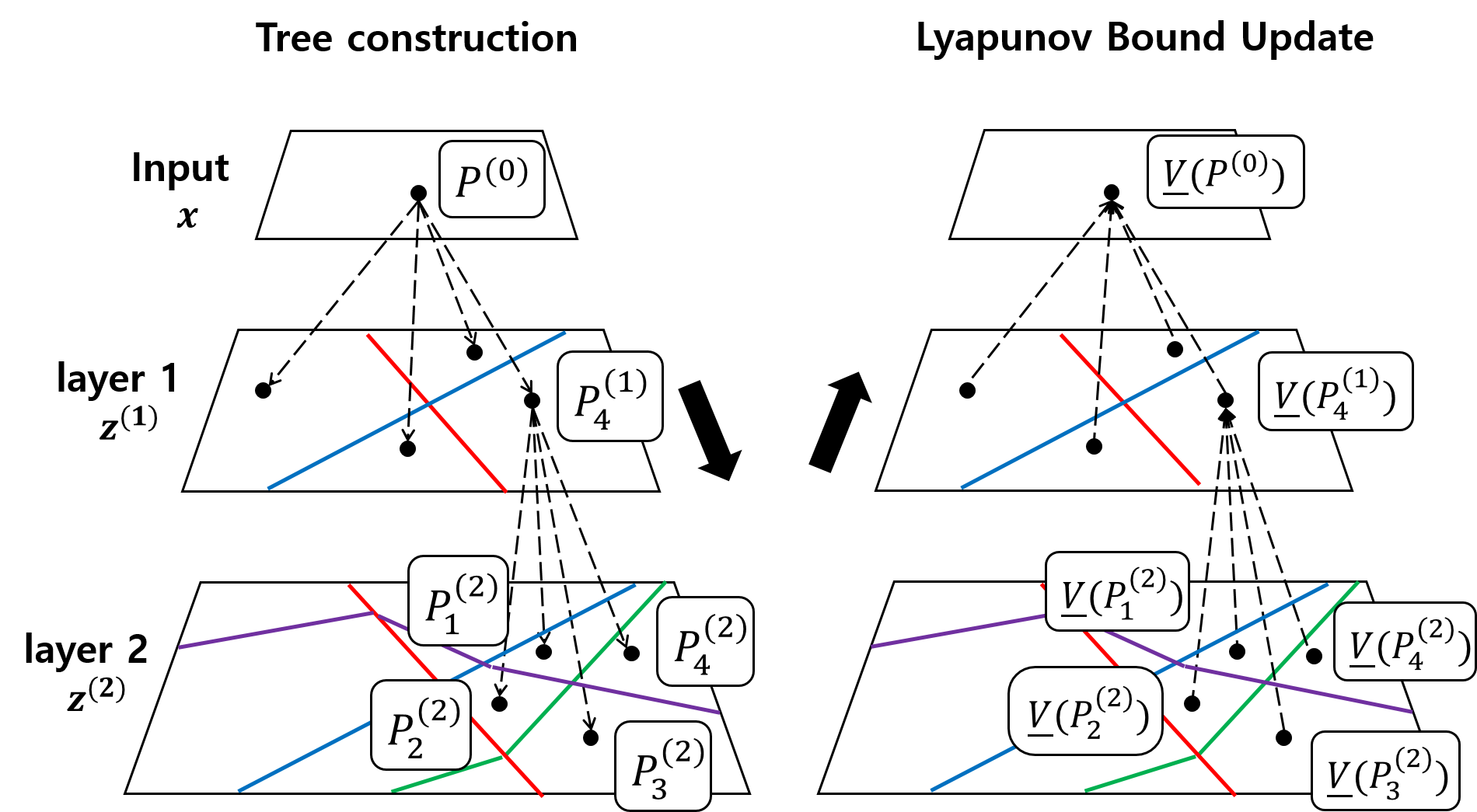}
    \caption{Description about (left) construction of the partition tree  and (right) updating lower bound of the Lyapunov value for each partition, for the NN with two hidden layers and two neurons for each layer. Colored lines indicate the hyperplanes generated by each neuron.}  
    \label{fig:pg}
\end{figure}

The partitions of the PWA NN are generated by divisions of the input space by each neuron at the breakpoints in the activation function. For the $j$-th neuron of the layer $l$, it creates the hyperplane over all partitions from the layer $l-1$, $P^{(l-1)}\in \mathcal{R}_{l-1}$, as 
\begin{align}
    H_{j}^{(l)} (P^{(l-1)}) = \{\x\in P^{(l-1)} | z_{pre,j}^{(l)}(\x) = 0 \}, \label{eq:neuron_hyperplane}
\end{align}
which can be computed with \eqref{eq:nn_hidden}. If the hyperplane divides the partition $P^{(l-1)}$, it creates new smaller partitions of the layer $l$. Thus, the procedure can be represented with tree structure, \textit{a partition tree}, as in \cite{tondel2002computation} and \cite{jouret2023safety}. The leaf nodes of the tree created from the layer $L-1$ indicates the partition set which is a convex polytope and the value is affine over the partition, which can be used to formulate the PWA NN in the form of a PWA function \eqref{eq:pwa}. 



In this subsection, we describe how a partition tree $\mathcal{T}$ is constructed from the PWA NN Lyapunov function. The procedure is illustrated in Fig. \ref{fig:pg} and the algorithm is described in Alg. \ref{alg:p_graph_construction}.
The procedure consists of two parts, \textit{extending the tree} and \textit{computing the lower bound of the Lyapunov function}. First, in order to extend tree for each layer, we compute the hyperplanes induced by the neurons from the layer. 
Since the generated hyperplanes depend on the partition from the previous layer, they are computed for each partition from the previous layer (lines 5-13). 
If the set of hyperplane $\mathcal{H}$ divides the partition $P$ into a  set of smaller partitions $\mathcal{P}'$ (line 10), it is added to the tree as children of the current partition (line 11). Following this process for every node in the tree and all layers, we can enumerate all partitions generated by the PWA NN. 
For every partition $P$ at the last layer, coefficients of the PWA NN  $C_{P}$ and $\mathbf{d}_{P}$ can be computed from \eqref{eq:nn_output}. 

After the tree extension is finished, we compute the maximum and minimum of the Lyapunov value on partitions from every layer. Starting from the leaf nodes, we compute the bounds of the Lyapunov value by solving LPs (line 19). Then, moving upward to its parent partition, each partition computes its bound by comparing the bounds of its children (line 21). For the partition $P$, the minimum Lyapunov value $\underline{V}(P)$ is  updated from its children $\{P'_{1},\cdots, P'_{N}\}$ following 
\begin{align}\label{eq:lyap_update}
        \underline{V}(P) =\min_{} \{ \underline{V}(P'_{1}), \cdots, \underline{V}(P'_{N})\}.     
\end{align}

Note that due to the reference symmetry of the system, we do not need to re-build the partition tree for different reference $\r$ and is built offline for the zero reference.

\begin{algorithm}[t] 
\caption{ Partition Tree Construction} \label{alg:p_graph_construction}
\textbf{Input:} Lyapunov Function $V_{NN}$, State Space $D$\\
\textbf{Output:} Partition Tree $\mathcal{T}$ 
\begin{algorithmic}[1] 
\State $R_{0}\leftarrow D$ (Initial Partition), $\mathcal{T}\leftarrow$ Initialize()
\State  \small{\texttt{/** Extend Partition Tree **/}}
\For{ $i=1,\cdots,L-1$} 
\State $\mathcal{R}_{i}\leftarrow\emptyset$ 
\For{$P \in \mathcal{R}_{i-1}$}
\State $\mathcal{H}\leftarrow \emptyset$
\For{$j=1,\cdots,w_{i}$} 
\State $\mathcal{H}\leftarrow \mathcal{H} \cup \{H_{j}^{(l)}(P)\}$
\EndFor
\State $\mathcal{P}'=\{P'_{1},\cdots,P'_{n(P,\mathcal{H})}\}$ $\leftarrow$ Divide $P$ with $\mathcal{H}$
\State Insert new nodes $\mathcal{P}'$ to $\mathcal{T}$ as children of $P$
\State $\mathcal{R}_{i}\leftarrow \mathcal{R}_{i} \cup \mathcal{P}'$
\EndFor
\EndFor
\State \small{\texttt{/** Lower Bound of the Lyapunov Value**/}}
\For{ $i=L-1,\cdots,1$} 
\For{$P \in \mathcal{R}_{i}$}
\If{$i=L-1$}
\State Compute $\underline{V}(P)$ with LP
\Else
\State Update $\underline{V}(P)$ with \eqref{eq:lyap_update}
\EndIf
\EndFor
\EndFor
\end{algorithmic}
\end{algorithm}

\subsection{Algorithmic Implementation}

In this subsection, we describe how to compute the maximal admissible Lyapunov level from the given reference $\r$ and the partition tree $\mathcal{T}$. First, we assume that we only consider constraint partitions $P_{c}$ such that $\max_{\x\in P_{c}}c(\x)>0$, meaning that at least a single inadmissible state exists. Otherwise the problem \eqref{prob:single_lp} is infeasible. 

The pseudocode for the algorithm is given in Alg. \ref{alg:invariant_level}. First, if the constraint is met at the given reference $c(\bar{\x}_{\r})=0$, it is true that $\Gamma^{*}=0$ from Lemma 6 of \cite{cotorruelo2021reference} (lines 2-4).
In order to find the maximal admissible Lyapunov level, we traverse the partition tree $\mathcal{T}$ in a bread-first manner. For every partition popped from the working set $W$ (line 7), if it is a leaf node, \eqref{prob:single_lp} is solved for each constraint partition $P_{c}\in \mathcal{P}_{c}$ and $\Gamma^{*}$ is updated (lines 12-21). Otherwise, its children partitions are inserted into the working set $W$ (line 23). 

Solving \eqref{eq:opt_problem} by traversing all nodes in the tree requires $| \mathcal{P}_{c}| \cdot | \mathcal{P}_{V}|$ number of LP subproblems. We design pruning strategies that discard LP subproblems which lead to  infeasible problem or unimportant results that do not help to find the optimal $\Gamma^{*}$. 

The first strategy (later denoted as PS1) is based on the fact that neurons from the first layer generate hyperplanes on the input space. If a hyperplane from the first layer $H$ has an empty intersection with the constraint partition $P_{c}$, every partitions $P_{V}$ which is on the other side of $P_{c}$ with respect to  $H$ automatically results in an infeasible LP, since $P_{V}\cap P_{c}=\emptyset$. Checking which side of the partition $P_{V}$ is located from $H$ can be done with the activation pattern. Therefore, we can skip the infeasible LPs (line 14-15) by gathering pairs of such inactive hyperplanes and constraint partitions. The procedure to find pairs of inactive hyperplanes for the constraint partition is described in Alg. \ref{alg:modify_pruning}, and it is executed prior to the traversal of the partition tree (line 5).
The pruning strategy generates additional $|\mathcal{P}_{c}|\cdot W_{1}$ number of LP subproblems.


The second strategy (later denoted as PS2) compares the lower bound of the Lyapunov value of the partition, which is computed from Alg. \ref{alg:p_graph_construction}, with the current estimate of the maximal admissible Lyapunov value (line 9). If the lower bound of the Lyapunov value is greater, the solutions of \ref{prob:single_lp} for leaf nodes derived from the partition are unnecessary since it would not change the current estimate of the optimal solution. Thus, these partitions are excluded from the on-going procedure. The two pruning strategies can be used both or only one strategy can be used based on the structure of the neural Lyapunov function.





\begin{algorithm}[t] 
\caption{Computation of Maximal Admissible Lyapunov Level } \label{alg:invariant_level}
\textbf{Input:} Lyapunov Network $V_{NN}$, Reference $\r$, Constraint Partition Set $\mathcal{P}_{c}$ \& Coefficient Set $\mathcal{C}_{c}$, Partition Tree $\mathcal{T}$\\
\textbf{Output:} Maximal Admissible Lyapunov Level $\Gamma^{*}$ 
\begin{algorithmic}[1] 
\State $\Gamma^{*}\leftarrow \infty$
\If{$c(\bar{\x}_{\r},\r)=0$}
\State \textbf{return} $\Gamma^{*}\leftarrow0$
\EndIf
\State $S_{1},\cdots,S_{|\mathcal{P}_{c}|} \leftarrow$ Alg. \ref{alg:modify_pruning} for $P_{c}\in \mathcal{P}_{c}$
\State $W\leftarrow \{\mathcal{T}\text{.root}\}$
\While{ $W$ is not empty} 
\State $R\leftarrow$ pop first element of $W$
\If{$\underline{V}(R) > \Gamma^{*}$}
\State \textbf{continue} 
\EndIf
\If{$R$ is leaf node}
\For{$k=1,\cdots, |\mathcal{P}_{c}|$}
\If{$\llambda^{(1)}(R)$ has match in $S_{k}$}
\State \textbf{continue}
\EndIf
\State $\Gamma\leftarrow$ Solve Problem 1. for $(R,\mathcal{P}_{c}(k))$ 
\If{$\Gamma < \Gamma^{*}$}
\State $\Gamma^{*} \leftarrow \Gamma$
\EndIf
\EndFor
\Else 
\State $W\leftarrow W \cup R.\text{children}$
\EndIf
\EndWhile
\end{algorithmic}
\end{algorithm}

Since the breadth-first search over the partition tree can visit all elements of the partition set $\mathcal{P}_{V}$ and since the LP subproblems discarded by the pruning strategies are either infeasible or do not change the estimate of the optimal solution, Alg. \ref{alg:invariant_level} can find the solution for \eqref{eq:opt_problem}.

\begin{remark}[Convex Polytope Constraint Set]
When the constraint set $K=\{\x: c(\x)\geq 0\}$ is a convex polytope, we can reduce the series of LP subproblems \eqref{prob:single_lp}  for the constraint partition set into a single LP, skipping iteration over $\mathcal{P}_{c}$. 
For the convex constraint set $K$ which can be formulated as $A_{c}\x \leq b_{c}$, solving multiple \eqref{eq:opt_problem} for each constraint partition can be replaced to solving the following single LP,
\begin{subequations}
\begin{align}
    \Gamma_{P_{V}}(\r)= \min &\hspace{2mm}  C_{V}(\x-\bar{\x}_{\r})+d_{V} \\
    \text{s.t.} \hspace{2mm} &\x \in P_{V}, \\
     \hspace{6mm} & A_{c}\x \leq b_{c}, \\
     \hspace{6mm} &C_{c}\x+d_{c} = 0 
\end{align}
\end{subequations}
which reduces the number of LP decreases by a factor of $|\mathcal{P}_{c}|$. 
It can be done similarly for the case where the exterior of the constraint set $K^{c}=D\setminus K$ is a convex polytope.     
\end{remark}




\begin{algorithm}[t] 
\caption{Finding Inactive Hyperplanes} \label{alg:modify_pruning}
\textbf{Input:} Hyperplane Set from the First Layer $\mathcal{H}$, Constraint Partition $P_{c}$
\\ \textbf{Output:} Set of neuron index and activation pattern $S$ 
\begin{algorithmic}[1]
\State $S\leftarrow \emptyset$
\For{$j=1,\cdots,w_{1}$}
\State $H\leftarrow \mathcal{H}(j)$ 
\If{$H \cap P_{c} = \emptyset$}
\State $S\leftarrow S \cup \{(j, \neg \lambda_{j}^{(1)}(P_{c})\}$
\EndIf
\EndFor
\end{algorithmic}
\end{algorithm}

\section{Data-Driven Estimation of the Maximal Admissible Lyapunov Level}

Even though the maximal admissible Lyapunov level $\Gamma^{*}$ can be computed from Alg. \ref{alg:invariant_level} and heuristics are developed to increase computational efficiency, it might not meet the requirement for real-time online control due to the increased number of partitions and larger network size. For faster inference, we develop a data-driven estimator $E_{NN}(\r):R\rightarrow\R$ that computes the admissible Lyapunov level  and a following training method that uses Alg. \ref{alg:invariant_level} as an oracle and makes the resulting estimator always return an admissible  Lyapunov level where its sublevel set is a CAPI set. 

To train the estimator which always returns an admissible Lyapunov level for given reference $\r \in \mathcal{R}$, we develop a counterexample-guided approach with a learner and a verifier. The verifier checks whether the current estimator is valid and it not, it generates counterexample. And this counterexample is added to the training set which the learner uses to train the estimator. 
\subsection{Verification}
First, we describe how the verification of the estimator is done. The verification process is done by solving the following optimization problem. 
\begin{problem}[Estimator Verification] \label{problem:verification}   
\begin{subequations}
\begin{align} \label{eq:verifier}
    \max_{\r \in \mathcal{R},\x \in \mathcal{X}} \hspace{3mm} &c(\x) \\
    s.t. \hspace{4mm} &V_{NN}(\x,\r) + s = \xi, \\
    &\xi = E_{NN}(\r), \\
    & s \geq 0,
\end{align}
\end{subequations}
\end{problem}
where $s\in \R$ is a slack variable. 
 The optimization problem finds the maximum violation from the sublevel set inferred from the estimator $\{\x \in \mathcal{X},\r\in\mathcal{R}: V_{NN}(\x,\r) \leq E_{NN}(\r)\}$. If the optimal value is positive, it is counter-example and if the optimal value is non-positive, the estimator is verified. Since the objective and constraints are all PWA function, the optimization problem can be transformed to mixed-integer linear program which exact solution can be found using solvers such as gurobi \cite{gurobi}. 



\subsection{Training}
The training process is described in Alg. \ref{alg:learn_estimator}. With the oracle $Q$ which computes the optimal admissible Lyapunov level from Alg. \ref{alg:invariant_level}, a pre-training dataset is generated by uniformly sampling references inside the domain $\mathcal{R}$ (line 2). This dataset is used to update the parameters of the estimator network $\boldsymbol{\theta}_{e}$ (line 3). Then, an iterative sequence of solving the verification problem \ref{problem:verification} and checking whether the optimal solution verifies the estimator follows. If the estimator is not verified, the counterexample $\r^{*}$ is added to the dataset and the parameters are updated (line 10). This process iterates until verification succeeds or the maximum number of iterations is reached. Finding conditions under which the algorithm always finds the solution is left as 
 future work. 

The loss function used for parameter updates is defined as the mean squared error between the return values from the estimator and the true maximal admissible Lyapunov level from the oracle,   $\mathcal{L} = \sum_{i=1}^{|D|} \| E_{NN}(\r_{i}) - Q(\r_{i}) \|$.

Training an additional estimator requires more offline computation, but the result from Alg. \ref{alg:learn_estimator} can always compute admissible Lyapunov level with cheap forward pass of NN. 
If the algorithm terminates with a small number of training sets, it might result in the estimator returning an underestimated value of the maximal admissible Lyapunov level. This might lead to conservativeness in controllers or planners using the estimator. In practice, to overcome the issue, we add a sufficient number of data points to the pre-training dataset. 

\begin{algorithm}[t] 
\caption{ Learning the Maximal Admissible Lyapunov Level Estimator}
\textbf{Input:} Constraint Function $c$, Oracle $Q$, Learning Rate $\alpha$\\
\textbf{Output:} Certified Estimator $E_{NN}$ 
\begin{algorithmic}[1] 
\State Initialize estimator parameters $\boldsymbol{\theta}_{e}$
\State Generate pre-training data $D=\{\r_{i},Q(\r_{i})\}_{i=1:n_{D}}$
\State Pre-train estimator $E_{NN}(\r;\boldsymbol{\theta}_{e})$
\For{$k=0,1,\cdots$}
\State $OPT,\r^{*}\leftarrow$Solve Prob. \ref{problem:verification}
\If {$OPT<0$}
\State \textbf{return} $E_{NN}(\r;\boldsymbol\theta_{e})$
\Else
\State $D\leftarrow D \cup \{\r^{*}, Q(\r^{*})\}$
\State $\boldsymbol{\theta}_{e}\leftarrow \boldsymbol{\theta}_{e} - \alpha \nabla_{\boldsymbol{\theta}_{e}}\mathcal{L}(D,\boldsymbol{\theta}_{e})$
\EndIf
\EndFor
\end{algorithmic} \label{alg:learn_estimator}
\end{algorithm}


\section{Evaluation} \label{sec:eval}

In this section, we evaluate the proposed approach to compute CAPI set. All numerical computations are performed on a Python environment with an Intel i9 CPU. We used Seidel's LP algorithm \cite{seidel1991small} to solve LP problems. The Lyapunov functions are parameterized by the NN with ReLU activation, which are trained by following \cite{wu2023neural} even though our algorithm is not restricted to a specific training method.  


\begin{figure}[t]
    \centering
    \begin{tabular}{cc}
      \subfloat[]{\includegraphics[height=0.38\linewidth]{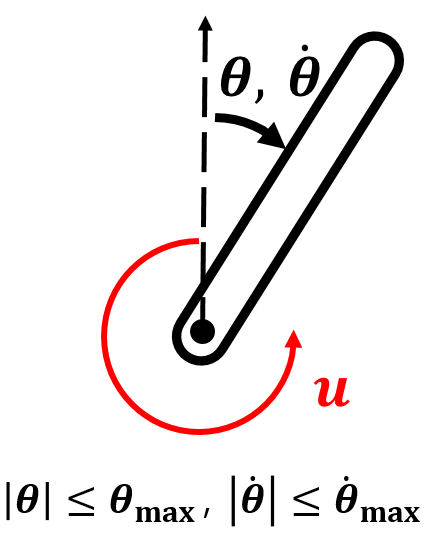}}
    &
      \subfloat[]{\includegraphics[height=0.38\linewidth]{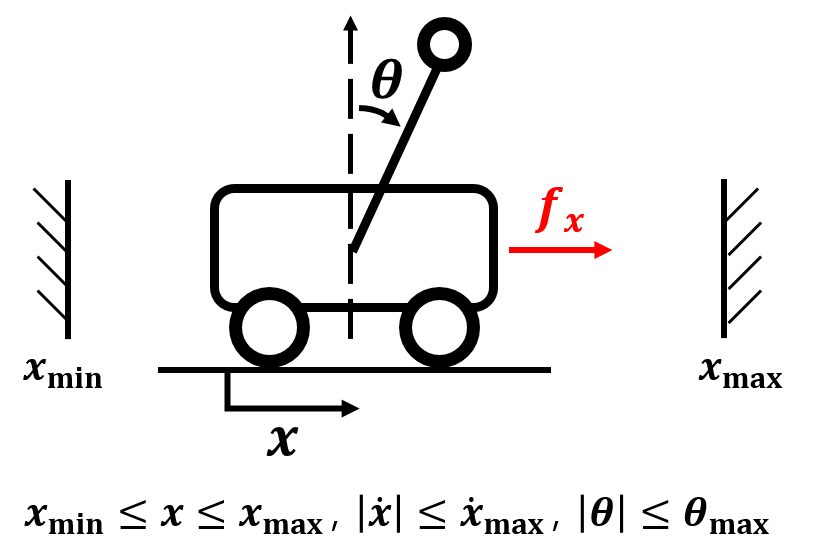}} 
    \end{tabular}
    \caption{Diagrams for (a) an inverted pendulum and (b) a cart-pole systems for evaluation of the proposed method with its constraints.}
    \label{fig:eval_diagrams}
\end{figure}

\begin{figure}[t]
    \centering
    \vspace{1mm}
        \begin{tabular}[t]{cc}
		  \subfloat[]{\includegraphics[width=0.475\linewidth]{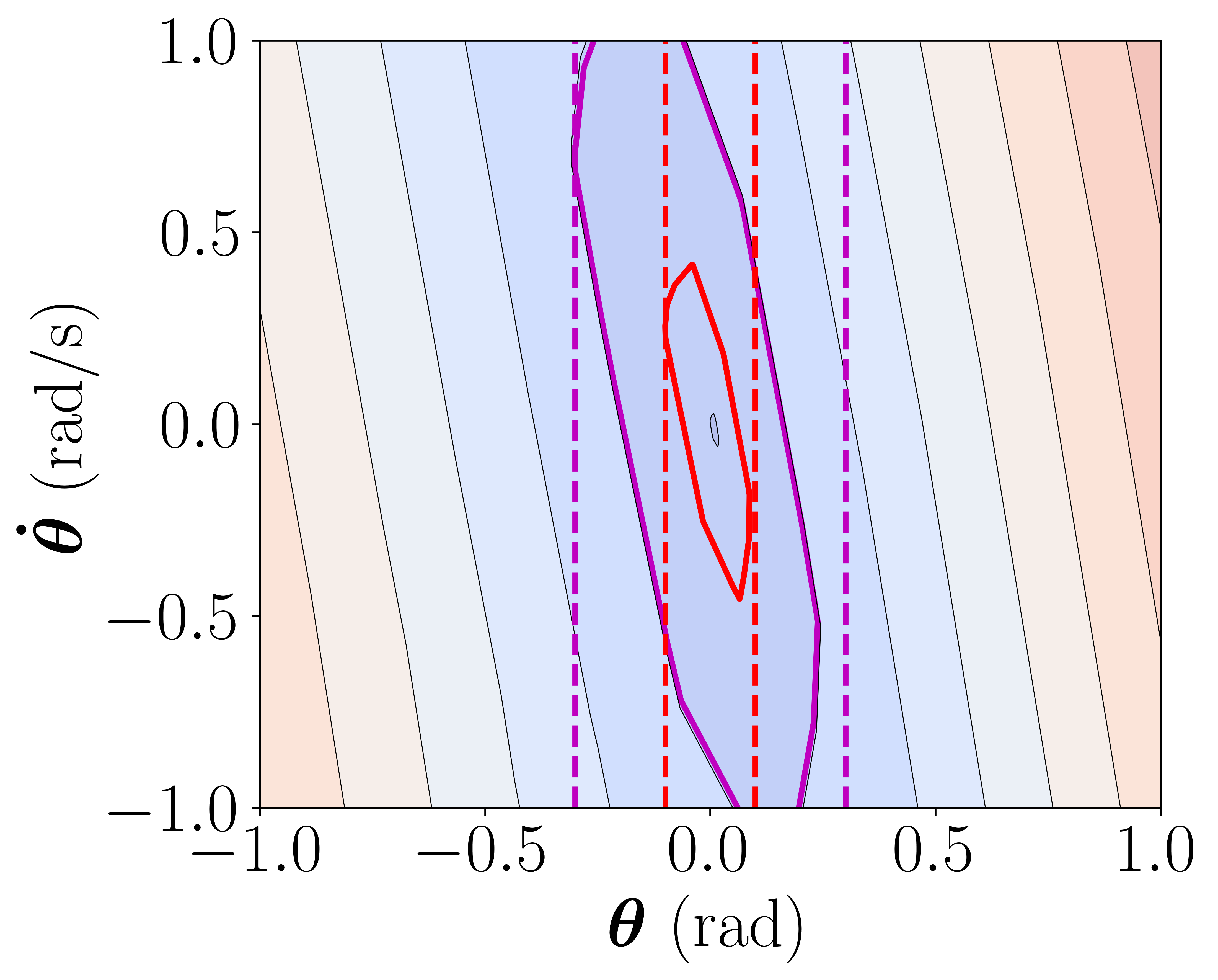}}
        & \hspace{-1mm}
		  \subfloat[]{\includegraphics[width=0.46\linewidth]{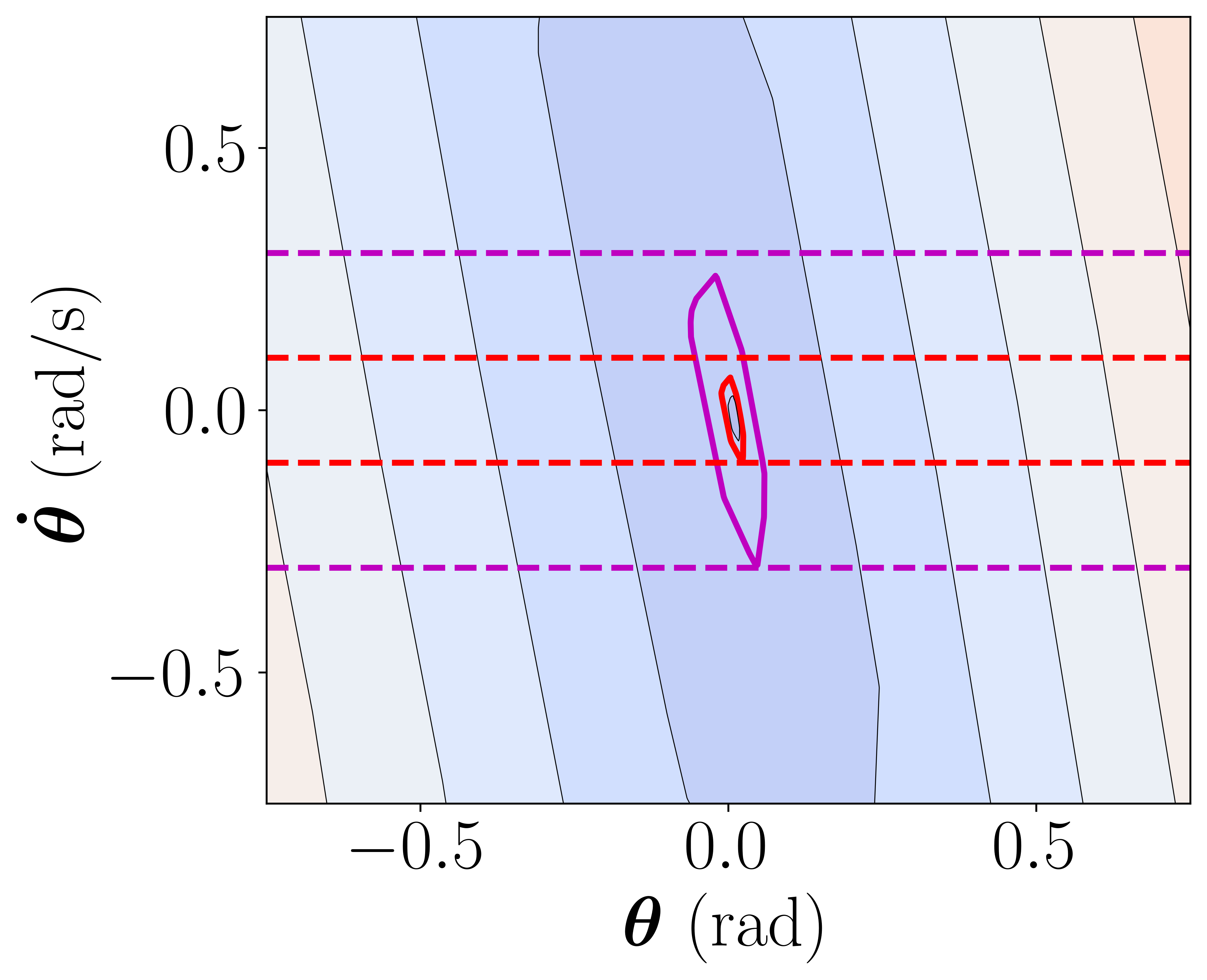}}
    \end{tabular}
    \caption{Results of computing maximal CAPI set for the invertend pendulum dynamics for constraints on bound of (a) $\theta$ and (b) $\dot{\theta}$.  Contour plot is the Lyapunov function value, red and purple dashed lines indicate the boundary of the admissible set, and bold lines indicate the surface of the resulting CAPI set.}
    \label{fig:inv_pend_result}
\end{figure}

\subsection{Case 1: Inverted Pendulum}
The first example is inverted pendulum dynamics with state space $\x=[\theta,\dot{\theta}]^{T}\in \R^{2}$ following, 
\begin{align}
    \ddot{\theta}= \frac{mgl\sin(\theta)+u-b\dot{\theta}}{ml^{2}}
\end{align} with gravity constant $g=9.81$, mass $m=0.15$, pendulum length $l=0.5$, and friction coefficient $b=0.1$. The dynamics is discretized with a time interval $\tau=0.05$,  and the input $u$ is bounded in $u\in[-6.0,6.0]$. The Lyapunov function has a  single hidden layer with 8 neurons, and the policy is $u=[-2.20, -0.638]^{T}\x$. The given constraints are either $|\theta| \leq \theta_{max}$ and $|\dot{\theta}| \leq \dot{\theta}_{max}$. The dynamics has the set of equilibrium state which is singleton $\{(\theta,\dot{\theta}):(0,0)\}$. 

The result is described in the two figures at Fig. \ref{fig:inv_pend_result} for different bounds in $\theta$ and $\dot{\theta}$ respectively. The maximal admissible Lyapunov level is found from Alg. \ref{alg:invariant_level}, and we can find that it can compute the maximal sublevel which meets the boundary of the admissible region. By differing $\theta_{max}$ and $\dot{\theta}_{max}$ from $0.05$ to $1.0$, we measure the computation time of each algorithm for each constraint and reported at Table \ref{tab:comp_time}. For computation of  Alg. \ref{alg:invariant_level}, we only used the second pruning strategy, PS2, since the Lyapunov network has  single hidden layer with small number of partitions.

\newcolumntype{M}[1]{>{\centering\arraybackslash}m{#1}}
\newcolumntype{N}{@{}m{0pt}@{}}


\subsection{Case 2: Cart-Pole}
The second example is a cart-pole system with state vector $\x=[x,\dot{x},\theta,\dot{\theta}]^{T}\in \R^{4}$ and the dynamics following 
\begin{subequations}  
\begin{align}
    \ddot{x} &= \frac{f_{x}+m_{p}\sin\theta (l\dot{\theta}^{2}-g\cos\theta)}{m_{c}+m_{p}\sin^{2}\theta},\\
    \ddot{\theta} &= \frac{-(f_{x}+m_{p}l\dot{\theta}^{2}\sin\theta)\cos\theta+(m_{c}+m_{p})g\sin\theta}{l(m_{c}+m_{p}\sin^{2}\theta)},
\end{align}
\end{subequations} 
with mass of the cart $m_{c}=1.0$, mass of the pole $m_{p}=0.1$, length of the pole $l=1.0$, gravity constant $g=9.81$, and discretized with $\tau=0.05$. The reference of the system is $r=x$ with corresponding equilibrium $\bar{\x}_{r}=[r,0,0,0]^{T}$. The given constraints are 
$x\in [x_{min},x_{max}]$, $|\theta| \leq \theta_{max}$, and $|\dot{x}| \leq \dot{x}_{max}$.
The Lyapunov function has two hidden layers with 12 neurons respectively, and zero bias terms. The control policy is given as $f_{x}(\x,r)=[ 1.09,  1.81, 34.6, 11.3]^{T}(\x-\bar{\x}_{\r})$. 


\begin{figure}
    \centering
    \vspace{2mm}
        \begin{tabular}[t]{cc}
		  \subfloat[]{\includegraphics[width=0.28\linewidth]{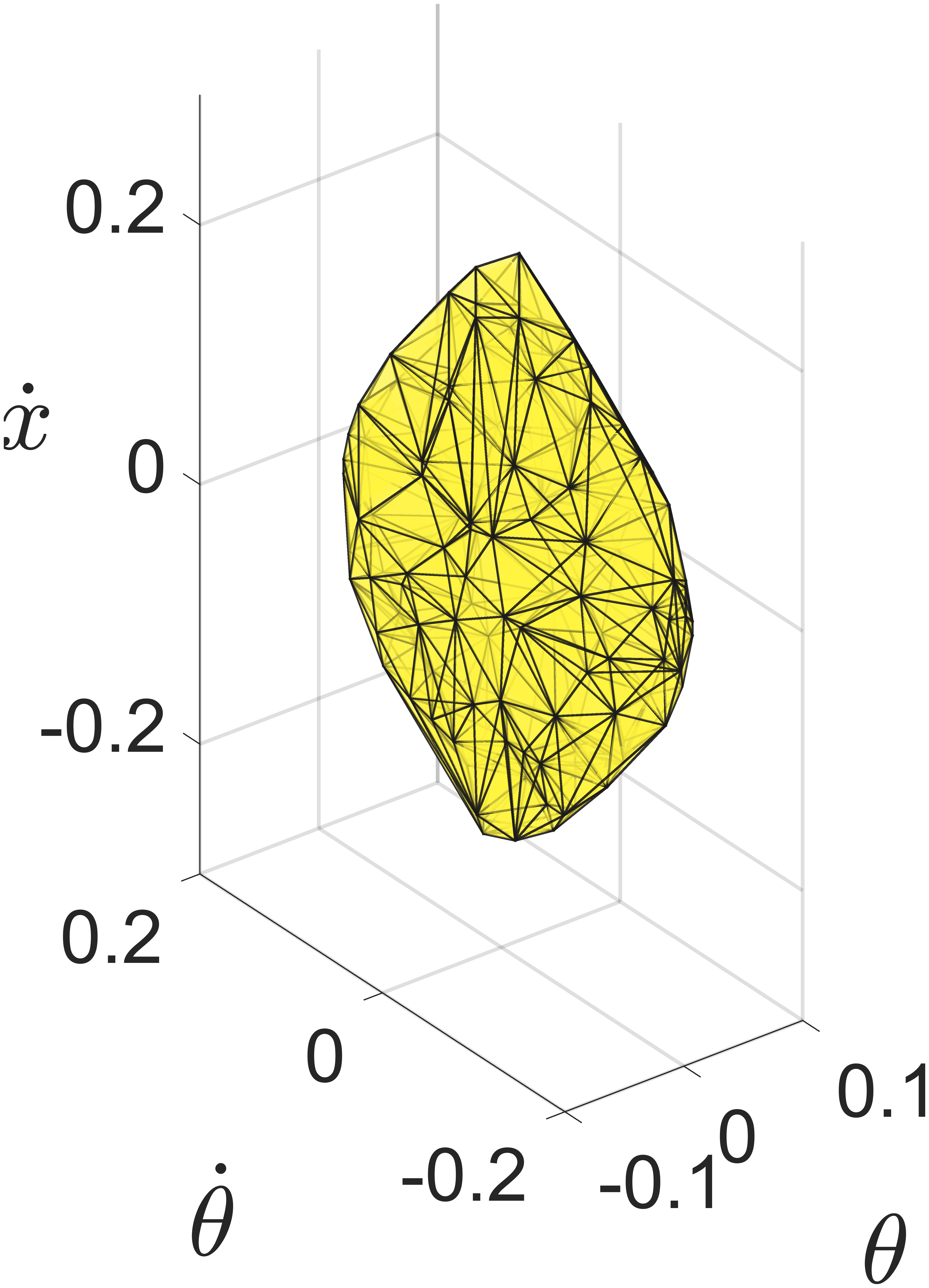}} 
        &
		  \subfloat[]{\includegraphics[width=0.6\linewidth]{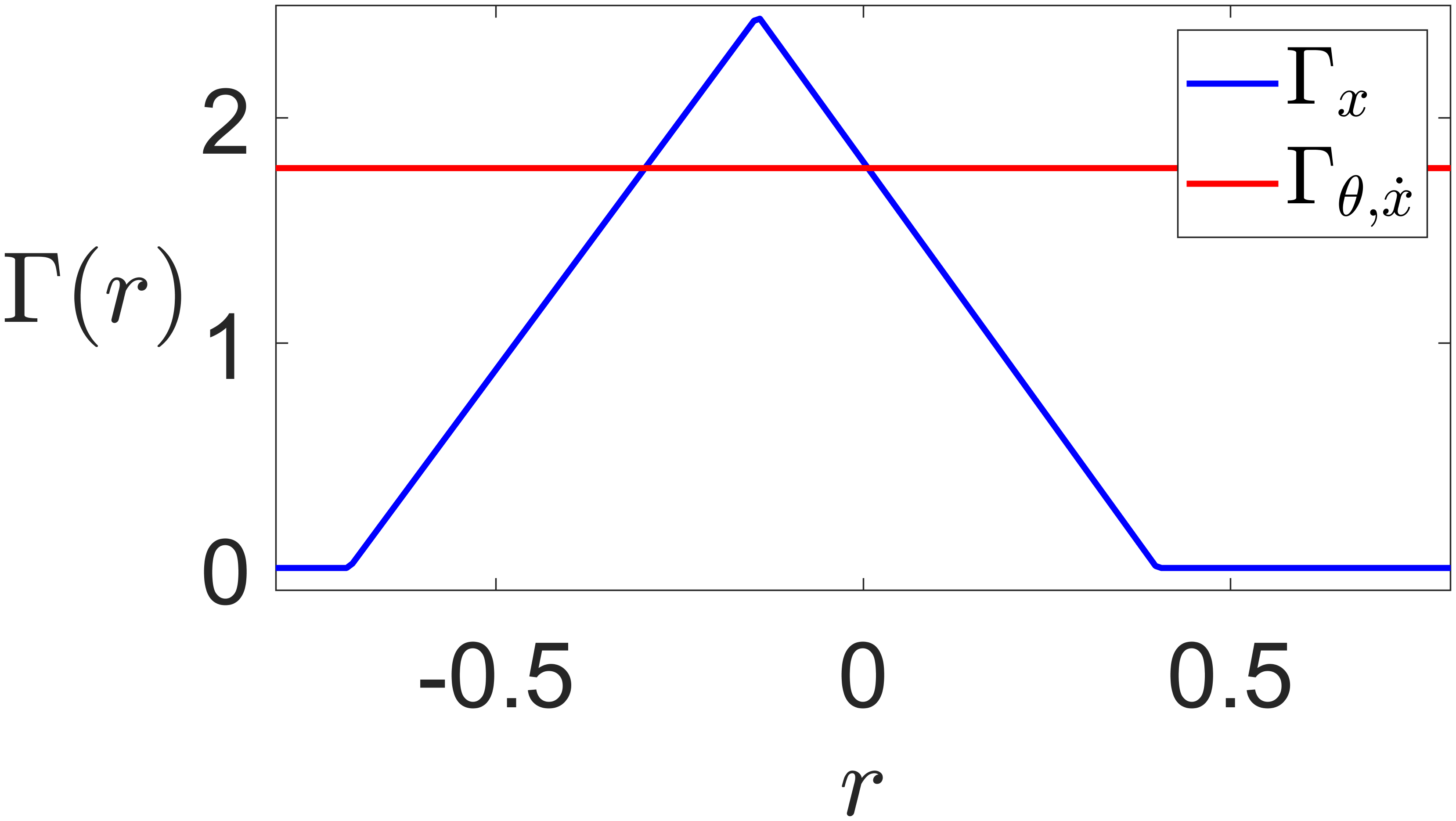}}
    \end{tabular}
    \caption{(a) Surface of the Lyapunov function level at $\Gamma_{\theta,\dot{x}}$, when $x=0$. (b) Plot for $\Gamma^{*}(r)$ where $\Gamma_{x}$ is computed with respect to the constraint $x\in [x_{min},x_{max}]$, and $\Gamma_{\theta,\dot{x}}$ is from the other constraints.}
    \label{fig:cartpole_result}
\end{figure}

Since constraints with respect to $\theta$ and $\dot{x}$ are independent of the reference, its maximal admissible Lyapunov level is also independent of the reference. For $\theta_{max}=0.1$ and $\dot{x}_{max}=0.3$, the maximal admissible Lyapunov level $\Gamma_{\theta,\dot{x}}$ is computed, and the sublevel set at $x=0$ is described in Fig. \ref{fig:cartpole_result} (a). For $x_{min}=-0.7$, $x_{max}=0.4$, the maximal admissible Lyapunov level $\Gamma_{x}(r)$ is computed following Alg. \ref{alg:invariant_level} and the resulting $\Gamma^{*}(r)$ can be found in Fig. \ref{fig:cartpole_result} (b). The computation time is measured for different values of $\dot{x}_{max}$, $\theta_{max}$, and $\x_{max}$, which are uniformly sampled from 0.05 to 1.0 and reported at Table \ref{tab:comp_time}. During computation of $\Gamma^{*}$ with Alg. \ref{alg:invariant_level}, we only used the first pruning strategy, PS1,  since the minimum Lyapunov value on every partitions are zero due to the zero bias terms of the Lyapunov network. 

As an application of computing a CAPI set, we designed an explicit reference governor (ERG) \cite{garone2015explicit} which generates an admissible reference $v$ from the desired reference $r$ with a feasibility guarantee of the closed-loop system. It updates the admissible reference $v$ by computing time derivative $\dot{v}$ as 
\begin{align}
    \dot{v} = \Delta(\x,v)\rho(r,v). \label{eq:erg}
\end{align}
Here, $\Delta$ is the dynamic safety margin (DSM), which is a scalar value that indicates the feasibility of the current state and the admissible reference $v$. $\rho$ is the navigation field (NF) which is a normalized vector field from the current admissible reference $v$ toward the desired reference $r$. 
The DSM can be computed as the difference between $\Gamma^{*}$ and the current Lyapunov value, as 
\begin{align}
    \Delta(\x,v)=\eta \cdot (\Gamma^{*}(v)-V(\x,v)) \label{eq:dsm}
\end{align}
where $\eta>0$ is a tuning parameter. $\Delta\geq0$ indicates that the current state $\x$ is inside the sublevel set of $\Gamma$, therefore the state $\x$ and the current reference $v$ is feasible. 

We design the ERG for the cart-pole dynamics with the same constraints, targeting a reference of $r = 0.399$ from the initial state $\x_{0} = [-0.4, 0, 0, 0]^{T}$. The NF is defined as $\rho(r,v) = 1$ if $r > v$, $\rho(r,v) = -1$ if $r < v$, and $\rho(r,v) = 0$ otherwise. The ERG, formulated in continuous time, was implemented in discrete time using the Euler method. The method for ensuring constraint satisfaction and goal convergence in discrete time is detailed in \cite{momani2024discrete}.

Since the ERG updates the admissible reference \eqref{eq:erg} at each control iteration, $\Gamma^{*}$ must be computed within the time interval. However, as shown in Table \ref{tab:comp_time}, Alg. \ref{alg:invariant_level}
 takes too long for real-time use. To solve this, we trained a fast estimator using Alg. \ref{alg:learn_estimator}, with a neural network (ReLU activations, two hidden layers of 8 and 4 neurons). This enables quick inference suitable for online control.




\begin{figure}[t]
    \centering
    \vspace{3mm}
   {\includegraphics[width=0.85\linewidth]{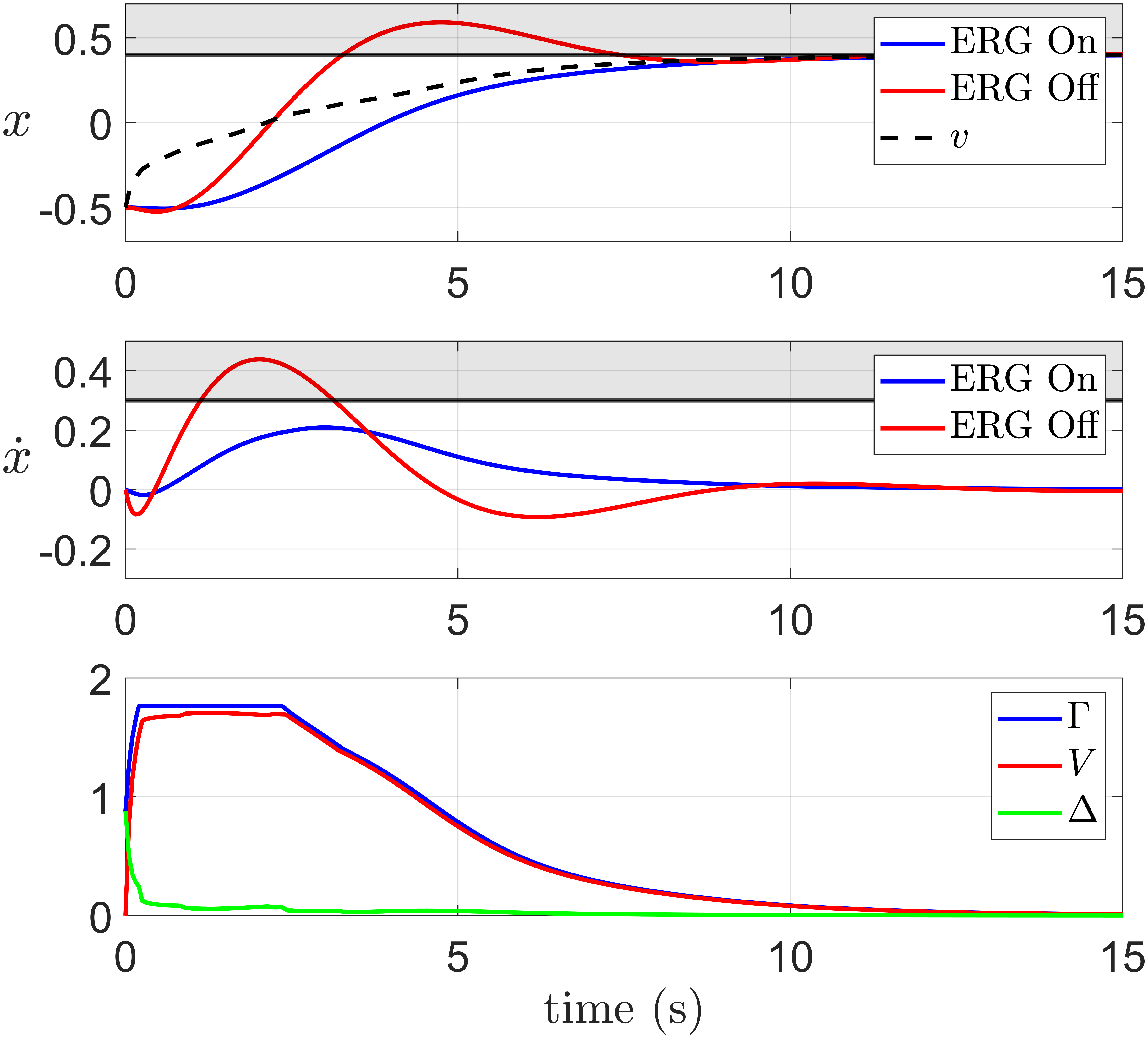}}
    \caption{Results from the ERG designed for the cart-pole system. With the admissible reference $v$ computed from the ERG, constraints are not violated for (a) $x$ and (b)  $\dot{x}$ states. (c) DSM $\Delta$ is computed as \eqref{eq:dsm} with $\eta=2$ and it is kept positive.}
    \label{fig:cartpole_erg}
\end{figure}

\begin{table}[t]
\renewcommand{\arraystretch}{2}
\footnotesize
    \centering
    \caption{\small{Numerical results of the evaluation on the inverted pendulum and the cart-pole dynamics. (WP: with pruning strategy, WOP: without pruning strategy)}}
    \begin{tabular}{|m{1.5cm}|m{2.0cm}|m{1.5cm}|m{1.4cm}|N} 
\hline
     \multicolumn{2}{|c|}{} & Inverted Pendulum  & Cart-pole &\\
  \hhline{|====|}
    \multirow{2}{*}{\parbox{1\linewidth}{Partition Tree}} &  Comp. Time (s) & 0.289 &  59.9 &\\
    \cline{2-4}
 & \# of Partitions & 32 & 1580  &\\
  \hhline{|====|}
    \multirow{2}{*}{\parbox{1\linewidth}{ \vspace{3mm}  Finding $\Gamma^{*}$ \\ Comp. Time }} & Alg 2. WP (ms) & \parbox{3cm}{mean: 0.496 \\ max:  0.725}  & \parbox{3cm}{mean: 21.4 \\ max: 33.8}  &\\[4pt]
    \cline{2-4}
    & Alg 2. WOP (ms) & \parbox{3cm}{mean: 0.732  \\ max:  0.943}  & \parbox{3cm}{mean: 26.0  \\ max:  34.6 }\\[4pt]
  \hhline{|====|}
    \multirow{2}{*}{\parbox{1\linewidth}{  Estimator \\ Comp. Time }} & Training (s) & - & 160 &\\
     \cline{2-4}
     & Inference ($\mu$s) & - & \parbox{3cm}{mean: 14.2 \\ max: 26.5} &\\[4pt]
\hline
    \end{tabular}
    \label{tab:comp_time}
\end{table}

The result from the ERG is described in Fig. \ref{fig:cartpole_erg}. In Fig. \ref{fig:cartpole_erg} (a), control without the ERG violates constraints and enters the unsafe region, while with the ERG, $x$ converges to the goal without crossing into unsafe areas, using $\Gamma^{*}$ from the neural Lyapunov function.
Fig. \ref{fig:cartpole_erg} (c) shows the temporal profile of $\Delta$, which decreases to zero near the safe set boundary but never drops below, ensuring safety.

\section{Conclusion}

Computing the constraint admissible positively invariant (CAPI) set is essential in constrained control and planning. Building on recent advances in neural Lyapunov functions, we propose a method to find the maximal CAPI set using PWA activations and constraints, simplifying the problem into smaller LP subproblems rather than relying on exhaustive mixed-integer programs. Additionally, we introduce a method to train an estimator for real-time computation of the maximal admissible Lyapunov level.

While this work uses a pre-trained Lyapunov function, future research will aim to simultaneously optimize CAPI sets and Lyapunov functions within a learning-based framework. Another promising direction is to reduce the reliance on Lyapunov analysis by leveraging contraction theory and incremental stability analysis \cite{forni2013differential}.

\addtolength{\textheight}{-12cm}   




\bibliographystyle{./bibtex/IEEEtran}
\bibliography{./bibtex/IEEEabrv, ./bibtex/mybibfile} 

\end{document}